\documentclass[10pt,letterpaper]{article}
\usepackage[english]{babel}
\usepackage{graphicx}

\newcommand{\alt}{\mathbin{\lower 3pt\hbox
   {$\rlap{\raise 5pt\hbox{$\char'074$}}\mathchar"7218$}}}
\newcommand{\agt}{\mathbin{\lower 3pt\hbox
   {$\rlap{\raise 5pt\hbox{$\char'076$}}\mathchar"7218$}}}

\begin{document}
\setcounter{footnote}{0}
\setcounter{equation}{0}
\setcounter{figure}{0}
\setcounter{table}{0}
\vspace*{5mm}

\begin{center}
{\large\bf Scaling for level statistics \\
from self-consistent theory of localization }

\vspace{4mm}
I. M. Suslov \\
P.L.Kapitza Institute for Physical Problems,
\\ 119334 Moscow, Russia \\
E-mail: suslov@kapitza.ras.ru
\\
\vspace{5mm}
\end{center}

\begin{center}
\begin{minipage}{135mm}
{\bf Abstract } \\
Accepting validity of self-consistent theory of localization by
Vollhardt and W${\rm {\ddot o}}$lfle, we derive the relations of
finite-size scaling  for different parameters characterizing the
level statistics. The obtained results are compared with the
extensive numerical material for space dimensions  $d=2,\,3,\,4$.
On the level of raw data, the results of numerical experiments are
compatible with the self-consistent theory, while the opposite
statements of the original papers are related with ambiguity of
interpretation
 and existence of small parameters of the
Ginzburg number type.
\end{minipage}
\end{center}
\vspace{5mm}

\twocolumn

%\vspace*{1.5mm}
%PACS 03.65-w, 05.50+q, 11.10.Hi, 71.23.An
%\vspace*{1.5mm}

%\newpage
%\vspace{6mm}
\begin{center}
{\bf 1. Introduction}
\end{center}

The present paper continues the series of publications
\cite{1,2,3} devoted to a theoretical analysis of numerical
algorithms used for investigation of the Anderson transition.
These studies are motivated by  contradiction of numerical data
(see a review article  \cite{4}) with self-consistent theory
by Vollhardt and W${\rm {\ddot o}}$lfle \cite{5,6}, which
reproduces the main body of theoretical results and
according to certain arguments \cite{7,8} gives the correct
critical behavior. In particular, the numerical results
are incompatible with existence of the upper critical
dimension $d_{c2}=4$, which is a rigorous consequence of
the Bogoliubov theorem \cite{9} on renormalizability of
 $\varphi^4$ theory \cite{1}.  Since numerical modelling
is carried out independently by different groups
[4,10-17], the presence of trivial mistakes is surely
 excluded; however, all numerical algorithms are
empirical and  not based on a serious theoretical
ground.

The object for the present investigation is the scaling for
level statistics  \cite{10}, which
currently  became one of
the most popular algorithms \cite{11}--\cite{15}. Its
comparative simplicity is related with the fact that it
deals only  with the spectrum of the matrix
Hamiltonians and does not require a calculation of
eigenfunctions or conductivity.

The distribution function $P(\omega)$ for a spacing $\omega$
between the nearest levels is conveniently treated in terms of
the variable
$$
s=\omega/\Delta\,, \qquad \Delta = 1/\nu_F L^d  \,,
\eqno(1)
$$
where  $\Delta=\langle\omega\rangle$ is the mean  level spacing
in a finite system having a form of the  $d$-dimensional
cube of size $L$; $\nu_F$ is the density of states at the
energy of interest
(like the Fermi level). According to \cite{10}, there are
three actual distributions:  Wigner--Dyson $(W)$, Poisson
$(P)$ and critical $(c)$ (Fig.\,1):
%three actual distributions:  Wigner--Dyson $P_W(s)$, Poisson
%$P_P(s)$ and critical $P_c(s)$ (Fig.\,1):
%
\begin{figure}
\centerline{\includegraphics[width=2.8 in]{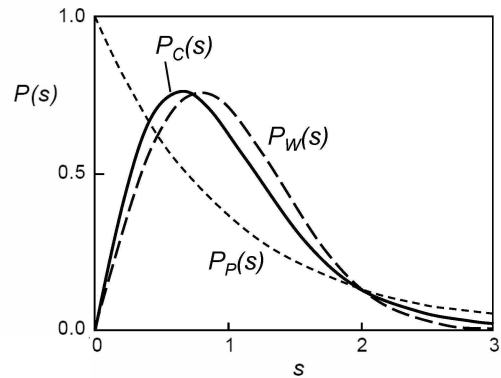}}
\caption{\footnotesize Distribution  $P(s)$ of the nearest
level spacing for Wigner-Dyson, Poisson and critical
statistics. Distributions $P_W(s)$ and $P_P(s)$
intersects in points  $s=0.473$ and
$s=2.002$.} \label{fig1} \end{figure}
$$
P_W(s)={\frac{\pi}{2}}\,s\,\exp
\left(-{\frac{\pi}{4}} s^2 \right)  \,, \qquad\qquad\qquad
\eqno(2)
$$
$$
P_P(s)=\exp \left(-s \right) \,, \qquad\qquad\qquad\qquad\qquad
\eqno(3)
$$
$$
P_c(s)=\left \{ \begin{array}{cc}
\sim s\,,& s\ll 1 \\
\sim \exp \left(-s/2\kappa \right)\,, & s\gg 1
\end{array} \right.\,,    \quad
\eqno(4)
$$
which are realized correspondingly in the metallic state,
the localization phase and the critical region. If the system is
in the critical point, then its level distribution coincides with
$P_c(s)$ independently of size $L$. With a small deviation
from the critical point, distribution  $P(s)$
changes slowly with  $L$ and tends to  $P_W(s)$ or $P_P(s)$
in the large $L$ limit. For a quantitative control of such
evolution one can consider the integral over
the large $s$ region,
$$
I(s_0)=\int_{s_0}^\infty \, P(s) ds \,,
\eqno(5)
$$
and introduce the scaling parameter
$$
\alpha(s_0)= \frac{I(s_0)-I_W(s_0)}{I_P(s_0)-I_W(s_0)}
 \,, \eqno(6)
$$
which changes from zero to unity with a crossover
%crossing over
from a metal to dielectric. If the scaling relation is
postulated,
$$
\alpha= F\left(L/\xi \right) \,, \eqno(7)
$$
then the critical behavior of the correlation length  $\xi$
can be extracted from the evolution of $\alpha$ under the change
of  $L$  \cite{10}.
Analogously, one can consider the integral over the small  $s$
region,
$$
\tilde I(s_0)=\int_0^{s_0} \, P(s) ds \,,
\eqno(8)
$$
and define the scaling parameter  $\tilde \alpha(s_0)$ analogously
to  (6), which  formally coincides with $\alpha(s_0)$ due to
relation  $\tilde I(s_0)=1- I(s_0)$. Practically,  definition (5)
is  traditionally used with the distinguished value $s_0=2.002$,
corresponding to the common intersection point of three
distributions (Fig.1), while definition (8) exploits a value
$s_0=0.473$ corresponding to the second intersection point of
$P_W(s)$ and $P_P(s)$.

Another variant of the scaling parameter is  coefficient
$A$ in the dependence
$$
I(s)=\exp\left( -As\right)
\eqno(9)
$$
which tends to a constant limit for large $s$;
 the scaling relation of type (7) can be
postulated for it. The more complicated versions of scaling
parameters were used in the cases  $d=2$ \cite{13} and
$d=4$ \cite{14}  (see  Secs.\,7,\,8).

The main questions are connected with scaling relations of
type (7), which cannot be justified for arbitrary
quantities, are certainly invalid in high dimensions and can be
essentially distorted by corrections to scaling. It is shown
below,  that self-consistent theory of localization
\cite{5,6} allows to establish the relations of type  (7) for all
introduced quantities, and the obtained scaling functions
can be compared with the extensive numerical material
\cite{10}--\cite{15}.  Analogously to
\cite{1}--\cite{3} it appears, that raw numerical data are
perfectly compatible with the  Vollhardt and W${\rm {\ddot
o}}$lfle theory, while  the opposite statements of
the corresponding
researchers are related with ambiguity of interpretation and
existence of small parameters of the Ginzburg number type.

\vspace{2mm}

\begin{center}
{\bf 2. Quasi-Gaussian Conception}
\end{center}

A calculation of the distribution function  $P(s)$ is practically
impossible for realistic models, and a theoretical analysis of
the algorithm looks rather
questionable. However, such analysis
becomes possible, if some roughening scheme is accepted. An example
of such a roughening is  the quasi-Gaussian conception
suggested by Altshuler et al   \cite{18}.

%Such
%a roughening is provided by the quasi-Gaussian conception
%suggested by Altshuler et al   \cite{18}.

Let $N$ be the number of levels in the interval  $E$ near
the  energy   $\epsilon_F$ (Fig.\,2); below  $\epsilon_F=0$
is accepted. If fluctuations of  $N$ are small, one can expect
a validity of the Gaussian distribution for them,
\begin{figure}
\centerline{\includegraphics[width=2.0 in]{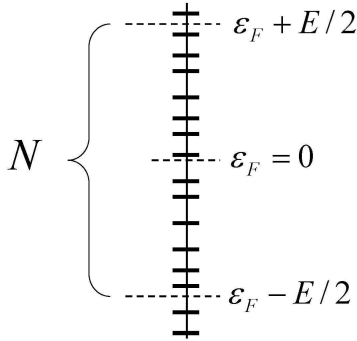}}
\caption{\footnotesize $N$ is the number of levels in the interval $E$.}
\label{fig2}
\end{figure}
$$
P(N)\sim \exp\left\{-\frac{(N-\langle N \rangle)^2}{2\sigma^2}
\right\}        \,,
\eqno(10)
$$
where  $\sigma^2$ depends on  $\langle N \rangle$. The
probability
of the event that there are no levels in the interval
%that not a single level is present in the interval
%of the event that none of levels belongs to the interval
$E$ is given by Eq.10  with  $N=0$. In terms of the introduced
quantities, it means that $\omega=s\Delta$ can take any
value greater than $E$; it corresponds to the integral  (5) with
 $s_0=E/\Delta$.  Taking into account a dependence of
$\sigma^2$ on $\langle N  \rangle=E/\Delta=s_0$, one has
$$
I(s_0)\sim \exp\left(-\frac{s_0^2}{2\sigma(s_0)^2}
\right)        \,.
\eqno(11)
$$
Since integration of $P(s)$ does not change the
form of the exponential in (2\,--\,4), one can reproduce it
by substitution
%setting
%accepting
$$
\sigma_W(s)^2 =2/\pi\,,\qquad
$$
$$
\sigma_P(s)^2 =s/2\,,\qquad
\eqno(12)
$$
$$
\sigma_c(s)^2 =\kappa s\,.  \qquad
$$
On the other hand, a direct calculation of the mean square
fluctuation
$$
\sigma_0^2= \langle N^2 \rangle  -\langle N \rangle^2
\eqno(13)
$$
gives
$$
(\sigma_0^2)_W =(2/\pi^2) \ln s\,,\qquad
$$
$$
(\sigma_0^2)_P = s\,,\qquad \qquad
\eqno(14)
$$
$$
(\sigma_0^2)_c = \kappa_0s\,,\qquad\qquad
$$
where the first expression is the result by Dyson \cite{19},
the second one corresponds to the Poisson distribution
\cite{20}, and the third  was suggested in \cite{18}
using the simple  scaling arguments  \cite{21} and
confirmed numerically in  \cite{11}. According to
\cite{11,14}
$$
\kappa_0 = 0.28\pm 0.03\,,\quad
\kappa = 0.26\pm 0.01\quad  (d=3)\,,
$$
$$
\kappa_0 = 0.45-0.50\,,\quad
\kappa \approx 0.36\qquad  (d=4)\,,
\eqno(15)
$$
i.e.  $\kappa$ and $\kappa_0$ are close but not identical. A
comparison of  (12) and (14) shows that  $\sigma^2$ and
$\sigma_0^2$ coincide in the order of magnitude aside from the
Wigner-Dyson case, where they differ by a logarithmic factor. The
latter is not surprising. Abundance of the Gaussian
distribution is a consequence of the central limit theorem, whose
derivation shows \cite{22}, that the Gaussian form is valid near
the maximum of distribution, while its tails remain not universal.
The given reasoning is valid in the certain interval of the $s$
values, which are sufficiently large for realization of the
exponential behavior in (2\,--\,4), but sufficiently small for a
crude validity of the Gaussian distribution  (10) in the vicinity
of  $N=0$. With any reasonable restrictions for  $s$,  one has $\ln
s \sim 1$ and the order-of-magnitude coincidence of $\sigma^2$ and
$\sigma_0^2$ is indeed  valid.
%takes place.
The two latter quantities vary in wide
limits and their difference in a slow function is
of a little consequence, so this function can be replaced by a
constant in
%the framework of
the accepted roughening scheme. As a result, an evolution of
distribution  $P(s)$ is mainly determined by the quantity
$\sigma_0^2$, which allows a theoretical description (Sec.\,3).

Substitution of (11) into (6) shows that for large  $s_0$
one can neglect
%the quantity
$I_W(s_0)$, so
$$
\alpha(s_0)= \exp\left\{-\frac{s_0^2}{2\sigma_P^2}
+\frac{s_0^2}{2\sigma^2} \right\} =
$$
$$
=\exp\left\{-s_0\frac{\sigma_P^2-\sigma^2}{\sigma^2}
 \right\}          \,,
\eqno(16)
$$
and  $\alpha(s_0)$ differs from zero only for $\sigma_P^2-\sigma^2
\ll \sigma_P^2$ and practically disappears in the Wigner-Dyson
range  $\sigma^2 \sim \sigma_W^2$. A comparison of (11) and (9)
shows that
$$
A= \frac{s}{2\sigma^2} = \frac{\sigma_P^2}{\sigma^2}  \,,
\eqno(17)
$$
so parameters  $\alpha(s_0)$ and  $A$ are determined by the
single combination  $\sigma^2/\sigma_P^2$; the same is true
for the more complicated scaling parameters (Secs.\,7,\,8).

\vspace{2mm}

\begin{center}
{\bf 3. Diagrammatic Approach}
\end{center}

A calculation of  $\sigma_0^2$ in the framework of the
diagrammatic technique was considered by Altshuler and Shklovskii
\cite{23}.  Having in mind the subsequent generalizations, we
discuss in details the selection principle of
%the  selection of
diagrams.

The number of levels  $N$ in the interval  $E$ is expressed
through the exact density of states  $\nu(\epsilon)$
in a finite system
$$
N= L^d\, \int\limits_{-E/2}^{E/2} \nu(\epsilon)
d\epsilon\,,\quad
\nu(\epsilon) = L^{-d} \sum\limits_{n}\,
\delta(\epsilon-\epsilon_n)\,,
\eqno(18)
$$
while its mean square fluctuation
$$
\sigma_0^2= L^{2d}\, \int\limits_{-E/2}^{E/2} d\epsilon_1
\int\limits_{-E/2}^{E/2} d\epsilon_2\,
K(\epsilon_1,\epsilon_2)
\eqno(19)
$$
is determined by the correlator
$$
K(\epsilon_1,\epsilon_2) =
\langle \nu(\epsilon_1)\nu(\epsilon_2) \rangle -
\langle \nu(\epsilon_1) \rangle
\langle \nu(\epsilon_2) \rangle .
\eqno(20)
$$
It is instructive to consider the quantity  $R(\omega)$,
which determines the probability to find
two arbitrary levels
at the distance $\omega$
(and not the nearest, as in the
case of $P(\omega)$); it is trivially connected with
$K(\epsilon_1,\epsilon_2)$
$$
R(\omega)=\frac{\langle \nu(E+\omega)\nu(E)
\rangle}{\langle \nu \rangle^2} =
\frac{K(E+\omega,E)}{\langle \nu \rangle^2}
+1
\eqno(21)
$$
(where  $\langle \nu(\epsilon) \rangle\equiv
\nu_F$ is assumed to be independent of $\epsilon$) and expressed through
the two-particle Green functions
$$
R(\omega)=\frac{\Delta}{2\pi^2 \nu_F}\,
{\rm Re}\frac{1}{L^{2d}} \, \sum\limits_{\bf k,q}
\left[\Phi^{RA}_{\bf kk}({\bf q})-
\Phi^{RR}_{\bf kk}({\bf q})  \right] \,.
\eqno(22)
$$
Here   $\Phi^{RA}_{\bf kk'}({\bf q})$ is a Fourier transform
of the quantity
$$
\Phi^{RA}({\bf r}_1,{\bf r}_2,{\bf r}_3,{\bf r}_4)=
 \left\langle G^{R}_{E+\omega}({\bf r}_1,{\bf r}_2)
    G^{A}_{E}({\bf r}_3,{\bf r}_4)  \right\rangle
\eqno(23)
$$
with the tree-momenta designations shown in
 Fig.\,3, and $\Phi^{RR}_{\bf kk'}({\bf q})$ is determined
 analogously. In terms of the vertex functions
 $\Gamma^{RA}_{\bf kk'}({\bf q})$ and
$\Gamma^{RR}_{\bf kk'}({\bf q})$ (Fig.\,3), one has
 $$
R(\omega)=1 + \frac{\Delta}{2\pi^2 \nu_F}\,
{\rm Re}\, L^{-2d} \, \sum\limits_{\bf k,q}
P_{\bf k}({\bf q})
\Gamma^{RA}_{\bf kk}({\bf q}) P _{\bf k}({\bf q}) \,,
\eqno(24)
$$
\begin{figure*}
\centerline{\includegraphics[width=5.4 in]{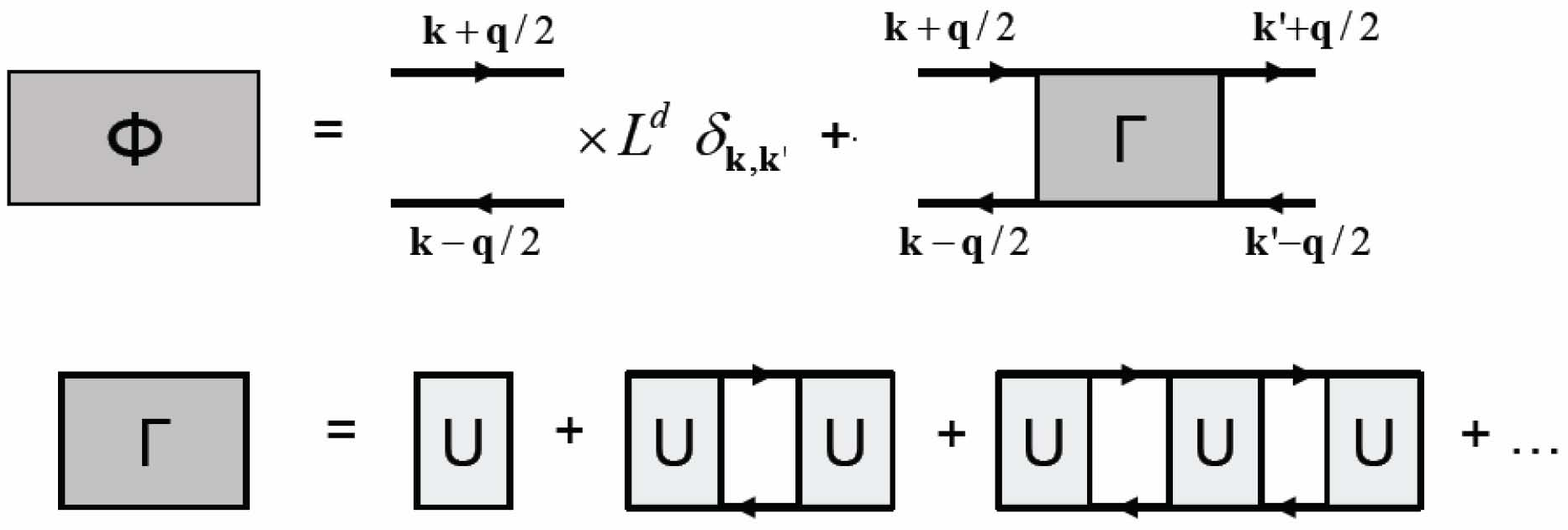}}
\caption{\footnotesize Relation of function $\Phi^{RA}_{\bf
kk'}({\bf q})$ with the full vertex $\Gamma^{RA}_{\bf kk'}({\bf q})$
and the irreducible vertex $U^{RA}_{\bf kk'}({\bf q})$. } \label{fig3}
\end{figure*}
where  $P_{\bf k}({\bf q}) =G^{R}_{{\bf k+q}/2} G^{A}_{{\bf
k-q}/2}$  and $\Gamma^{RR}_{\bf kk'}({\bf q})$ is omitted, since it
  gives no contribution
due to the absence of the diffusion poles  (see below).
The crucial point is the presence of the factor
$\Delta=1/\nu_F L^d$ before the sum over momenta in
Eq.24. If vertex $\Gamma^{RA}_{\bf kk'}({\bf q})$ is
regular, then the usual rule
for the change of summation by integration
$$
L^{-d} \, \sum\limits_{\bf k} \,\ldots
\longrightarrow  \int \frac{d^dk}{(2\pi)^d}\,\ldots
$$
gives  a finite expression multiplied by  $\Delta$, which
disappears in the thermodynamic limit.  In fact, vertex
$\Gamma^{RA}_{\bf kk'}({\bf q})$ contains the singular
contributions related with the diffusion poles, the so called
"diffusons" and "cooperons" (Fig.\,4,\,a,\,b),
\begin{figure*}
\centerline{\includegraphics[width=5.4 in]{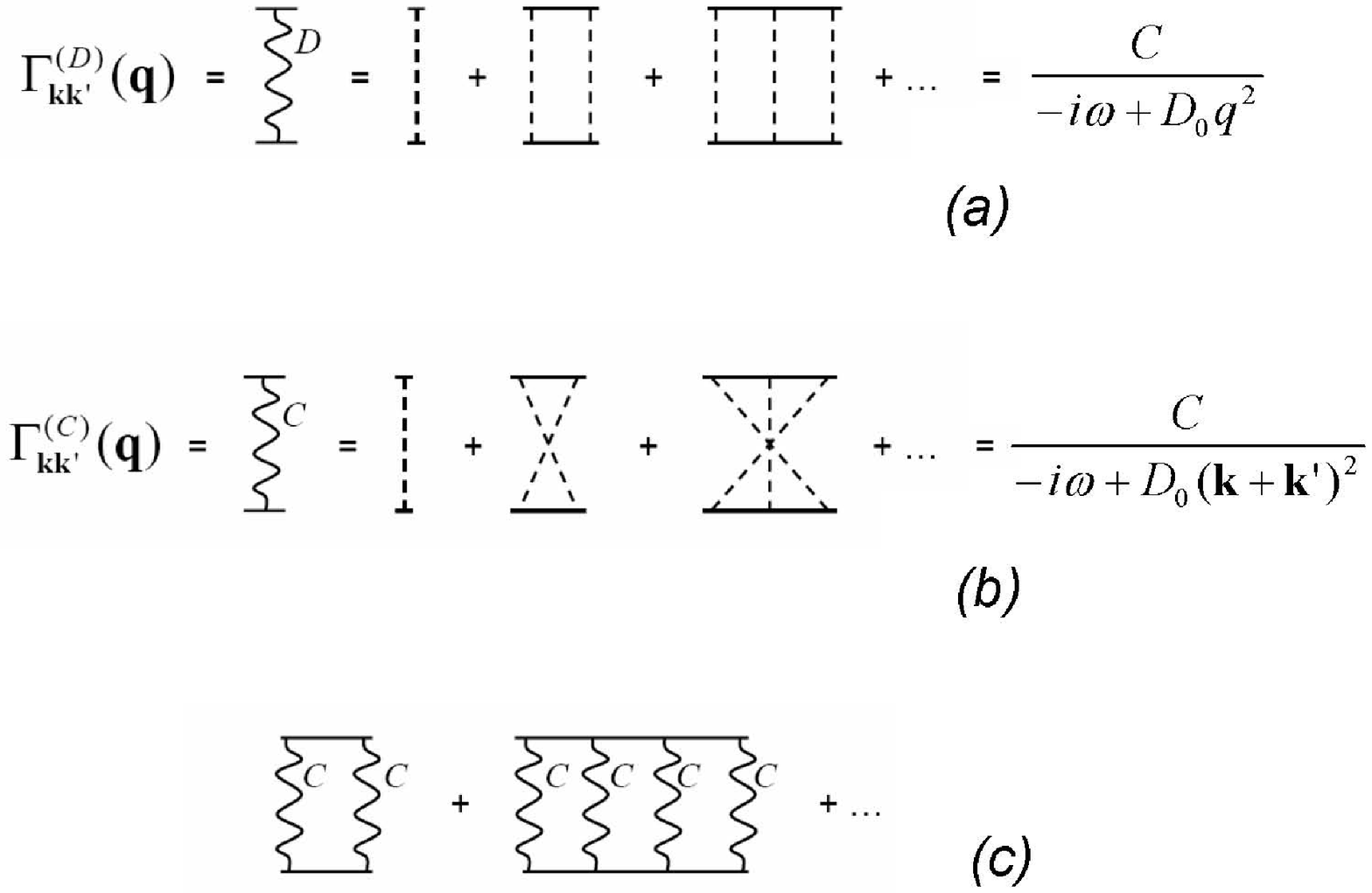}}
\caption{\footnotesize Definitions of the "diffuson" (a),
the "cooperon" (b), and the cooperon ladder (c). }
\label{fig4} \end{figure*}
which give singularities $1/\omega$ for certain values of momenta.
Fixation of the momentum at one value  (instead of summing) gives
factor  $ L^{-d}\propto \Delta $; if fixation of $(n-1)$ momenta
allows to nullify the momentum parts in  $n$ diffusion
denominators, then contribution $\Delta^n/\omega^n=1/s^n$ appears
in Eq.24, which remains finite in terms of variable $s$ when the
thermodynamic limit is taken. The simplest diagram in possession
of such a property is the two-cooperon one\,\footnote{\,It was
considered firstly by Bulaevskii and Sadovskii \cite{23a} and then
extensively used in \cite{23}.} (Fig.\,4,\,c)
$$
\Gamma^{CC}_{\bf kk'}({\bf q}) \sim
L^{-d} \, \sum\limits_{{\bf k}_1}
\frac{1}{-i\omega+D_0({\bf k+k_1})^2}\,\cdot
$$
$$
\cdot\,
P_{{\bf k}_1}({\bf q})                 \,
\frac{1}{-i\omega+D_0({\bf k}_1+{\bf k'})^2} 
\eqno(25)
$$
($D_0$ is a classical diffusion constant).  Since only the
vertex
with ${\bf k}={\bf k'}$ enters in Eq.24, then fixation of
 momentum ${\bf k}_1$  at value  $-{\bf k}$ nullifies the
momentum parts of two diffusion denominators and gives
contribution $1/s^2$ into  $R(s)$; the same contribution is given
by the diagram obtained from the two-cooperon one by reversing the
lower $G$-line\,\footnote{\,Factor 2 related with a possibility to
reverse the lower $G$-line is taken into account below in summing
the cooperon ladder.}, so
$$
R(s)=1-\frac{1}{\pi^2 s^2}\,,
\qquad
s =\omega/\Delta\,,
\eqno(26)
$$
which is a beginning of expansion over  $1/s$. Contributions
 $1/s^{2n}$ arise, in particular, from the ladder diagrams
containing $2n$ cooperons  (Fig.\,4,\,c). A summation  of
all such contributions should reproduce the Efetov result
 \cite{24} ($x=\pi s$):
$$
R(x)=1-\frac{\sin^2x}{x^2} -
\left(\frac{\sin x}{x}\right)'
\int_1^\infty \frac{\sin xt}{t}\,dt\,=
$$
$$
=\left \{ \begin{array}{cc}  \frac{\pi}{6}x\,,& x\ll 1 \\
1-\frac{1}{x^2} +\frac{1+\cos^2 x}{x^4}\,, & x\gg 1
\end{array} \right.\,,
 \eqno(27)
 $$
which corresponds to the Wigner-Dyson statistics. It is
interesting that a summation of the cooperon ladder  (Fig.\,4,\,c)
gives the result
$$
R(x)=\frac{1}{\pi}\,\int_{-\infty}^\infty
\frac{dt}{\sqrt{1+t^2} \sqrt{1+t^2+4 x^{-2}}  }
 \eqno(28)
 $$
which reasonably approximates  (27) (Fig.\,5).
\begin{figure}
\centerline{\includegraphics[width=2.5in]{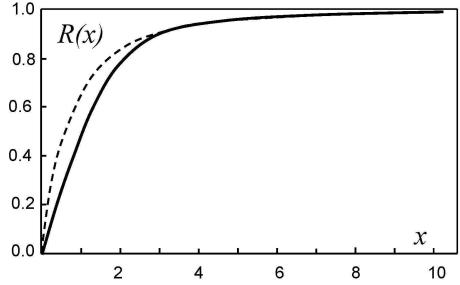}}
\caption{\footnotesize Comparison of the exact Efetov result
(solid line) and a contribution of the cooperon ladder (dashed
line). } \label{fig5}
\end{figure}
In its improvement, the main difficulty is related with
reproducing the weak oscillations, which are practically invisible
in Fig.\,5; the latter have the non-pertur- bative character and can
be obtained only if the factorial divergency of the perturbation
series is taken into account and the proper summation procedure is
used \cite{25,26}.

The analogue of the result (26) for the correlator
$K(\epsilon_1,\epsilon_2)$ has a form \cite{23}
$$
K(\epsilon_1,\epsilon_2) = \frac{1}{\pi^2 L^{2d}} \,
{\rm Re} \left( \frac{1}{-i\omega +\gamma}\right)^2\,,
\eqno(29)
$$
where  $\omega=\epsilon_1-\epsilon_2$ and attenuation
$\gamma$ is added, related with inelastic processes or
the openness of the system\,\footnote{\,If  imaginary
increments  $\pm i 0$ in the definitions of  $G^R$ and
$G^A$ are changed by $\pm i \gamma/2$, then replacement
 $-i\omega \to -i\omega +\gamma$ occurs in all diffusion
denominators \cite{2}.}. Substitution of (29)
 into (19) leads to expression\,\footnote{\,At first glance, the result
 (30) looks strange: expression  (29)  is localized at
 $|\omega|\alt \,\gamma$ and should give contribution
 $1/\gamma$ when integrated over  $\omega$, transforming
 to $E/\gamma$
%in the result of
after the second integration in
  (19). In fact, the integral over  $\omega$ in the infinite
  limits is zero and becomes finite only due to
  a restriction of   the integration domain; it leads to
  contributions $1/\epsilon_1$ and $1/(E-\epsilon_1)$,
  transforming to logarithms after integration over
  $\epsilon_1$.   } \cite{23}
$$
\sigma_0^2 = \frac{1}{\pi^2}
\ln  \frac{E^2+\gamma^2}{\gamma^2}\,,
\eqno(30)
$$
which coincides with Dyson's result (14) at $\gamma\sim \Delta$.
The latter fact has a following explanation. If a sufficiently
large attenuation $\gamma$ is artificially introduced, then
the two-cooperon contribution (29) is the main term of the
expansion in $\Delta/\gamma$, and Eq.30 is substantiated.
Dyson's result (14) refers to the closed systems and
implies $\gamma=0$. However, the condition of validity for (29)
allows to diminish  $\gamma$ only till a value of the
order of $\Delta$; fortunately, the dependence on $\gamma$
is practically absent for $\gamma\alt
\Delta$\,\footnote{\,It is clear from the fact that
a value of $R(s)$  at  $s=0$ can be obtained from Eq.26 at
$s\sim 1$.} and result (30) is matched with Dyson's
one. Below we use the same reasoning in the more
complicated case (Sec.4).

If a contribution to sum (25) is not restricted by the term
${\bf k}_1={-\bf k}$, but values of ${\bf k}_1$  close to
${-\bf k}$ are taken into account, then the
following result is 
obtained instead of  (29) \cite{23}:
$$
K(\epsilon_1,\epsilon_2) = \frac{1}{\pi^2 L^{2d}} \,
{\rm Re} \sum\limits_{\bf q} \left( \frac{1}{-i\omega
+\gamma+D_0 q^2}\right)^2\,.
\eqno(31)
$$
The restriction by the term ${\bf q}=0$ is justified for
$E\ll D_0/L^2$, while in the opposite case one can come
%change
from summation to integration and obtain
for  $E\gg\gamma$  \cite{23}:
%
%summation can be replaced by integration, leading to the
%result in the case $E\gg\gamma$  \cite{23}:
%
$$
\sigma_0^2 = \frac{1}{\pi^2 }
 \sum\limits_{\bf q} \ln \left[1+ \frac{E^2}
 {(\gamma+D_0 q^2)^2}\right]\,=
 a_d \left( \frac{L}{L_E}\right)^d,
 $$
 $$
a_d=\frac{K_d}{\pi d \sin(\pi d/4)}\,,
\eqno(32)
$$
where $L_E=(D_0/E)^{1/2}$ is the diffusion length over the
time $1/E$ and $K_d=\left[2^{d-1}\pi^{d/2}\Gamma(d/2) \right]^{-1}$
is the surface of the unit $d$-dimensional sphere, divided
by $(2\pi)^d$.

\begin{center}
{\bf 4. Application of Self-Consistent Theory}
\end{center}

The next step was made by Kuchinskii and Sadov- skii \cite{27}.
Results  (30,\,32)  are valid in the deep of the metallic phase,
and one can try to extend the region of their applicability,
replacing $D_0$ in (31) by the exact
diffusion coefficient  $D(\omega,q)$ \cite{27}
in the spirit of self-consistent theory of
localization \cite{5,6}.  Such
approach can be motivated by the following reasoning. The
irreducible vertex $U^{RA}$ (Fig.\,3) contains the
diffusion pole\,\footnote{\,The possibility to neglect
the spatial dispersion of $D(\omega,q)$ is justified in \cite{8}.}
$$
U^{RA}_{{\bf k} {\bf k}^\prime} ({\bf q}) =
U_{{\bf k} {\bf k}^\prime}^{reg}
({\bf q}) + \frac{F ({\bf k}, {\bf k}^\prime, {\bf q})}{-i \omega
 + D (\omega)({\bf k} + {\bf
k}^\prime)^2} \, \eqno(33)
$$
with the observable diffusion coefficient  $D(\omega)$.
Instead of the two-cooperon diagram  (Fig.\,4\,c), one can
consider the diagram with two blocks $U$ (Fig.\,3), which
%is main
dominates
in the metallic phase and under certain conditions (see below)
preserves domination in the  general
case\,\footnote{\,The diagrams with odd number of blocks  $U$
%have the additional smallness in
are suppressed by
parameter  $E/\gamma_e$,
where the elastic damping $\gamma_e$ has the order of the
bandwidth in the critical region. In terms of the $U$-blocks,
all diagrams are of the ladder type, and in this sense the
cooperon ladder  (Fig.\,4,\,c) corresponds to a summation of the
most singular contributions. The diagram with two  $U$-blocks is
the first term of this sequence, while the higher order diagrams
are discussed below.  }.  In the vicinity of the pole, one can
put  ${\bf k}'={-\bf k}$ in the function $F({\bf k},{\bf k}',{\bf
q})$ and  its role reduces to the additional factor $k_\sigma$
 after integration over ${\bf k},{\bf q}$ in  (24):
$$
K(\epsilon_1,\epsilon_2) =
\frac{k_{\sigma }\nu_F^2} {\pi^2 } {\rm Re} \sum\limits_{\bf q}
\left[ \frac{\Delta}{-i\omega +D(\omega) q^2}\right]^2\,.
\eqno(34)
$$
Factor $k_\sigma$ is a slow
function of a distance to the transition, which we
replace by a constant in  correspondence with the accepted
roughening scheme (Sec.\,2).

According to  \cite{2}, in a close finite system the diffusion
coefficient has a localization character
$$
D(\omega) =(-i\omega) \xi_{0D}^2\,,
\eqno(35)
$$
where $\xi_{0D}$ is the correlation length of a
 finite system considered as quasi-zero-dimensional.
Inelastic damping  $\gamma$ can be introduced by replacement
$-i\omega \to -i\omega +\gamma$, which is made
simultaneously in the term $-i\omega$ and in $D(\omega)$
 \cite{2}. Then
$$
K(\epsilon_1,\epsilon_2) =
\frac{k_{\sigma }\nu_F^2} {\pi^2 }\,  {\rm Re} \,
 \frac{\Delta^2}{\left(-i\omega +\gamma\right)^2}
\,F(\xi_{0D}/L)\,,
\eqno(36)
$$
where function $F(x)$ is defined as
$$
F(x)= \sum\limits_{\bf s}
\left[ \frac{1}{1 + (2\pi x {\bf s})^2}\right]^2\,
\eqno(37)
$$
and has the asymptotic behavior
$$
F(x)=\left \{ \begin{array}{cc} 1+O(1/x^4), & x\gg 1 \\
\tilde c_d/x^d\,, & x\ll 1
\end{array} \right.\,;
 \eqno(38)
 $$
Here  ${\bf s}=(s_1,\ldots, s_d)$ is a vector with integer
components  $s_i=0,\pm 1,\pm 2\ldots$ and $\tilde c_d=\pi
K_d(1-d/2)/2\sin(\pi d/2)$). Substitution of (36) into  (19) gives
$$
\sigma_0^2 = \frac{k_\sigma}{\pi^2}
\ln  \frac{E^2+\gamma^2}{\gamma^2}\,F(\xi_{0D}/L)\,
\eqno(39)
$$
instead of (30). We need an approximation providing
a correct description in the region  $\omega\sim \gamma$,
which plays an essential role in the integration
over $\epsilon_1,\,\epsilon_2$  (see Footnote 4),
%in  (19)
and where (36) is the main term of the expansion in
$\Delta/\gamma$. An example of the ladder diagrams
(Fig.\,4,\,c) shows that there exist contributions
$$
\left[ \frac{\Delta^2}{\gamma^2} \,F(\xi_{0D}/L) \right]^n
\eqno(40)
$$
with all  $n$, so the minimal  $\gamma$ providing a validity
of (36) is determined by  the condition
$$
\frac{\gamma_{min}^2}{\Delta^2} \,\sim \,F(\xi_{0D}/L)
\,
\eqno(41)
$$
and the inelastic damping cannot be diminished below
this quantity. Since a dependence on  $\gamma$ is practically
absent for  $\gamma\alt \gamma_{min}$   (see below), a value of
(36) at  $\gamma=0$  can be estimated setting $\gamma\sim
 \gamma_{min}$.  In proceeding to (39) one should take account of
 the $\omega$ dependence for $\xi_{0D}$ (Sec.\,5), which
  effectively adds contribution  $\sim E^2$ to the quantity
 $\gamma^2$ in the course of integration (19);
 hence, one should set
 $$
 \gamma^2=k_1 E^2 + k_2\gamma_{min}^2  \,,
 \eqno(42)
 $$
 where $k_1$ and $k_2$ are slowly varying functions and  can be
 approximated by constants. As a result,
 we have
$$
\sigma_0^2 = \frac{k_\sigma}{\pi^2} \,F(\xi_{0D}/L)
\ln  \frac{s^2+k_1 s^2 +k_2\,F(\xi_{0D}/L)}
{k_1 s^2 +k_2\,F(\xi_{0D}/L)}\,.
\eqno(43)
$$
In moving to the deep of the localized phase,  function
 $F(\xi_{0D}/L)$ grows to infinity and $\sigma_0^2$
% saturates by
tends to a constant, which accepts the Poisson value
$\sigma_P^2=s$ for  the choice  $k_2=k_\sigma s/\pi^2$;
so
$$
\frac{\sigma_0^2}{\sigma_P^2}
 = u \ln  \frac{1+k_1  +u}{k_1 +u}\,, \qquad
u=  \frac{k_\sigma}{\pi^2 s} \,F(1/z) \,,
$$
$$
z=L/\xi_{0D} \,.
 \eqno(44)
$$
Since  $\xi_{0D}/L$ is a function of  $\xi/L$ \cite{2},
the scaling relation of type (7) is established for the
quantity $\sigma_0^2/\sigma_P^2$.

Let  discuss the sense of relation (42) and a dependence on
$\gamma$ in the region $\gamma\alt\gamma_{min}$. A physical
interpretation of the result (32) is as follows: the system is
divided into quasi-independent blocks of size $L_E$  \cite{23} and
the nontrivial properties of $\sigma_0^2$ are formed at the scale
$L_E$, while for the larger scales there is addition of variances
as for independent random quantities.  The openness of each block
provides the diffusion attenuation $\gamma_D=D/L_E^2=E$ of its
eigenstates, with inelastic damping $\gamma$ added to it; they are
combined by the law of squares, since technically it involves an
estimate of $Re(-i\omega+\gamma)\sim (\omega^2+\gamma^2)^{1/2}$ at
$\omega\sim E$ (Sec.\,5).  Inelastic damping $\gamma$ is
inessential in the background of $\gamma_D$ under condition
$\gamma\alt E$. It will be clear below (Sec.\,5), that
$\gamma_{min}\sim E$ in the critical region and $\gamma_{min}\ll
E$ in the metallic one, so a dependence on $\gamma$ is absent in
both regions for $\gamma\alt\gamma_{min}$. In the localized
regime, the scale $L_E$ reduces to $\xi$ and the condition $E\ll
\Delta_\xi$ is fulfilled, where $\Delta_\xi$ is the level spacing
for a block of size  $\xi$. Under such condition, one can easily
estimate the probability $p_n$ for existence of
%that there are
$n$ levels in the interval $E$ for such
a block: $p_0\approx1-E/\Delta_\xi$, $p_1= E/\Delta_\xi$, $p_{n\ge
2}\approx 0$, so $\langle N \rangle \approx E/\Delta_\xi$,
$\langle N^2 \rangle \approx E/\Delta_\xi$ and $\sigma_0^2$
is close to the Poisson value independently of the actual
level statistics. Attenuation $\gamma$ can be considered as a
result of the random process, which provides the scattering
of each level near its average value; then independence of
statistics means independence of  $\gamma$. We see that  a
weak dependence on  $\gamma$ under condition  $\gamma\alt
\gamma_{min}$ takes place in all cases.

\begin{center}
{\bf 5. Scaling for Dynamical Conductivity and Dependence
on  $\omega$}
\end{center}

In the previous section, we  assumed implicitly   that the
quantity $\omega$ is sufficiently small. This assumption is not
valid in the general case, and the $\omega$ dependence
needs an additional study.

In a closed finite system, the diffusion coefficient has a
localization
 behavior  (35). In the passage to open systems, one
should make a replacement  $-i\omega \to -i\omega +\gamma$, and
the diffusion coefficient accepts a finite value $\gamma
\xi_{0D}^2$ in the static limit, leading to a finite conductance
$g_L$. The scaling relations for  $g_L$ and $\xi_{0D}$ were
derived in \cite{2} and have a form
$$
g_L=H_T\left(\frac{L}{\xi_{0D}}\right) \,, \quad
\pm c_d \left(\frac{L}{\xi} \right)^{d-2}
= H \left(\frac{L}{\xi_{0D}} \right) \,,
\eqno(45)
$$
where $c_d=\pi K_d/|2\sin(\pi d /2)|$ and functions
$H(z)$,  $H_T(z)$ have the asymptotic behavior
$$
H(z)=
\left \{ \begin{array}{cc}
\displaystyle{1/z^2}\,,& \qquad z\ll 1 \\
-c_d z^{d-2}\,,& \qquad z\gg 1
\end{array} \right.  \,,
$$
$$
H_T(z)=
\left \{ \begin{array}{cc}
\displaystyle{1/z^2}\,,& \qquad z\ll 1 \\
\sim e^{-z},& \qquad z\gg 1
\end{array} \right.  \,.
\eqno(46)
$$
Attenuation $\gamma_0$, arising due to the openness of
the system, is determined by relation
$$
\frac{\gamma_0}{\Delta} = z^2
H_T\left(z\right) \,, \qquad z=L/\xi_{0D} \,,
\eqno(47)
$$
so the ratio  $\gamma_0/\Delta$ is equal to unity in the
metallic phase,  a somewhat less  in the critical region
and  exponentially small in the localized state. Inelastic
damping $\gamma$, which we introduce for validity of formulas,
 is typically much greater and  $\gamma_0$ is
inessential in its background. The above relations
are valid in the limit of infinitesimal frequency and
need reconsideration for finite $\omega$.

The self-consistent equation of the  Vollhardt and W${\rm
{\ddot o}}$lfle theory can be written in the  form
\cite{1}
$$
\frac{{\cal E}^2}{W^2} =
\frac{D(\omega)}{D_{min}} + \Lambda^{2-d} \int\limits_0^\Lambda
\frac{d^dq}{(2\pi)^d} \,\frac{1}{[-i\omega/D(\omega)] + q^2} \,
\eqno(48)
$$
where ${\cal E}$ is the energy of the bandwidth order, $W$ is the
random potential amplitude, $\Lambda$ is ultraviolet cut-off,
$D_{min}$ is a characteristic scale of the diffusion constant,
corresponding to the Mott minimal conductivity $\sigma_{min}$, and
the limits of integration are indicated for the modulus of ${ q}$.

For finite  $L$, equation(48) accepts the following
form in the closed and open systems \cite{2}
$$
\frac{{\cal E}^2}{W^2} =  \frac{(-i\omega) \xi_{0D}^2}{D_{min}}
+\Lambda^{2-d} \cdot\frac{1}{L^{d}} \sum^{(c)}_{\rm q}
\,\frac{1}{q^2+m^2}  \,, \eqno(49)
$$
$$
\frac{{\cal E}^2}{W^2} = \frac{D_L(\omega)}{D_{min}} +
\Lambda^{2-d} \cdot\frac{1}{L^{d}} \sum^{(o)}_{\rm q}
\,\frac{1}{q^2+m^2}  \,,
\eqno(50)
$$
where $ m^{-1}= \xi_{0D}$. Symbols  $(c)$ and $(o)$
mark the allowed values of the momentum, corresponding
to the closed and open systems:  the main point is existence
of the term with ${\bf q}=0$ in the former case and its absence
in the latter \cite{2}. The first equation determines
$ \xi_{0D}$, while the difference of equations defines the
diffusion coefficient $D_L(\omega)$. Introducing the
dimensionless conductance $g_L(\omega)=h\nu_F D_L(\omega) L^{d-2}$
and producing transformations described in \cite{2}, one
obtains

$$
g_L= \frac{p}{z^2}+ H_T\left(z\right) \,,\quad
\pm c_d \left(L/\xi\right)^{d-2} =
\frac{p}{z^2}+ H \left(z \right) \,,
$$
$$
 p=(-i\omega+\gamma)/\Delta\,,\qquad  z =L/\xi_{0D}    \,,
\eqno(51)
$$
where inelastic attenuation  $\gamma$ is added.  Now the
quantity $\xi_{0D}$ depends on $\omega$ and its modulus
(at  $\gamma=0$) is usually denoted as $L_\omega$;
excluding $p$, we have the scaling for dynamical conductivity
$$
g_L(\omega)= F\left(L/\xi, L/L_\omega\right) \,,
$$
discussed by Shapiro and Abrahams \cite{29,30}. Equations
(51) transfer into (45) under condition
$|-i\omega+\gamma|\ll \Delta$, while the opposite case is
actual.

For $|p|\gg 1$, the large  $z$ region is of the main interest
where the second asymptotics (46) is valid for $H(z)$,
while $H_T(z)$ is exponentially small:
$$
g_L= \frac{p}{z^2} \,,\qquad
\pm c_d \left(L/\xi\right)^{d-2} =
\frac{p}{z^2}-c_d z^{d-2} \,.
\eqno(52)
$$
The localized regime takes place for $z\gg |p|^{1/d}$, where
$$
\xi_{0D}(\omega)=\xi_{0D}(0)\,,\quad
g_L(\omega)= \frac{-i\omega +\gamma}{\Delta} g_L(0)\,,
 \eqno(53)
$$
and $\xi_{0D}$ does not depend on frequency, so proceeding from
(36) to (39) in Sec.\,4 is substantiated; the quantities
$\xi_{0D}(0)$ and $g_L(0)$ are determined by Eqs.52 with $p=1$.
If $z\ll |p|^{1/d}$,  the metallic regime is realized,  where
$$
\xi_{0D}(\omega)=\left(\frac{-i\omega +\gamma}{\Delta}
\right)^{-1/2} \xi_{0D}(0)\,,
$$
$$
g_L(\omega)=g_L(0) \,,
\eqno(54)
$$
and the diffusion constant  $D$ is frequency-independent;
hence, the calculation by  Altshuler and Shklovskii  is adequate
and equation (32) is valid with the replacement of  $D_0$ by $D$.
In the critical region ($z\sim |p|^{1/d}$) both quantities are
$\omega$-dependent,
$$
\xi_{0D}(\omega)\sim\left({-i\omega +\gamma} \right)^{-1/d}
\,,
$$
$$
g_L(\omega)\sim\left({-i\omega +\gamma} \right)^{(d-2)/d}
 \,, \eqno(55)
 $$
so neither (39) nor (32)  is correct.

Substituting  (53--55) into (36) and using the second
asymptotics (38) for  $F(x)$, one can write
all three results in the unique form:
$$
K(\epsilon_1,\epsilon_2) =
\frac{k_{\sigma } \tilde c_d} {\pi^2 L^{2d} }\,  {\rm Re} \,
 \frac{1}{\left(-i\omega +\gamma\right)^2} \,\cdot
 $$
 $$\cdot\,
\left(\frac{-i\omega+\gamma}{\Delta} \right)^\beta
\left[\frac{L}{\xi_{0D}(0)} \right]^d\,,
\eqno(56)
$$
where the exponent  $\beta$ accepts values $0,\,1,\,d/2$ in the
localized phase, critical region and metallic state
correspondingly. Equation (56) can be considered as the
interpolation formula for the whole range of parameters, if  $\beta$
is understood as a slowly varying function. Substituting (56) into
(19) and integrating, one has
$$
\sigma_0^2 =
\frac{2 k_{\sigma } \tilde c_d} {\pi^2 }\,  {\rm Re} \,
 \frac{(\gamma+iE)^\beta -\gamma^\beta}
 {\beta(1-\beta)\Delta^\beta}
\left[\frac{L}{\xi_{0D}(0)} \right]^d\,.
\eqno(57)
$$
For $E\agt \gamma$, the right hand side of (57) is determined by
the term ${\rm Re} \,(\gamma+iE)^\beta \sim
(\gamma^2+E^2)^{\beta/2}$ and the same
result  by the order of magnitude
 follows from the expression\,\footnote{\,Condition $E\agt
\gamma$ is violated in the localized phase, but in this case
there is no dependence on the quantity $p$ and the character of
approximation for it has no significance.  }
$$
K(\epsilon_1,\epsilon_2) =
\frac{k_{\sigma } \tilde c_d} {\pi^2 L^{2d} }\,  {\rm Re} \,
 \frac{1}{\left(-i\omega +\gamma\right)^2} \,\cdot
 $$
 $$\cdot\,
\left(\frac{\gamma^2+E^2}{\Delta^2} \right)^{\beta/2}
\left[\frac{L}{\xi_{0D}(0)} \right]^d\,.
\eqno(58)
$$
It is easy to see that one can use Eq.36 with $\xi_{0D}$
independent of $\omega$, if combination $-i\omega+\gamma$ in
(52) is replaced by a quantity of the order
$(\gamma^2+E^2)^{1/2}$; since $\gamma\sim \gamma_{min}$,
it justifies representation (42) for the effective attenuation.

As a result, the second equation (52) accepts the form
$$
\pm c_d \left(L/\xi\right)^{d-2} =
\frac{s(k_1+u)^{1/2}}{z^2}-c_d z^{d-2} \,,
$$
$$ z=L/\xi_{0D}
\,,
\eqno(59)
$$
and together with (44) determines $\sigma_0^2$ as a function
of $L/\xi$. In the critical region, one has
$u\sim 1$ and  $\gamma_{min}$ appears to be of the order
of  $E$.

\begin{center}
{\bf 6. Three-Dimensional Case}
\end{center}

\begin{center}
{\bf 6.1. Scaling for  $\sigma_0^2$}
\end{center}

For large $s$,  we can use  the second asymptotics (38)
for  $F(1/z)$, make a replacement $u\to k_1 u$ and
exclude $z$, reducing  (44), (59) to the form
$$
\frac{\sigma_0^2}{\sigma_P^2}
 =k_1 u \ln  \frac{1+k_1  +k_1 u}{k_1 + k_1 u}\,,\qquad
$$
$$
\pm \left(\frac{L}{s^{1/d}\xi }\right)^{d-2} =
\frac{ (1+u)^{1/2} -Bu}{u^{2/d}} \,.\qquad
\eqno(60)
$$
We have changed the common scale of $\xi$, in order to have the
unit coefficient in the left hand side of the second equation,
and introduced the parameter  $B= \pi^2 c_d  k_1^{1/2}/(k_\sigma
\tilde c_d)$. Equations (60) are valid for dimensions  $2<d<4$
and in the parametric form determine the scaling
$$
\frac{\sigma_0^2}{\sigma_P^2}
 =F_\sigma \left(\frac{L}{s^{1/d}\xi }\right) \,,
\eqno(61)
$$
so the quantities  $L/\xi$ and  $s$ enter
only in the certain combination. Exactly such scaling was
discovered in numerical experiments \cite{11}.

We can make the proper choice of parameters  $k_1$ and
$k_\sigma$, in order  to reproduce the
correct results in the metallic phase and at the critical point.
Noticing that the scale $L_E$ coincide with $\xi_{0D}$ for
$p=s$, we have $\xi_{0D}=k_1^{-1/4} L_E$ from Eq.59 in the small
$z$ region; then  Eq.44 gives
$$
\sigma_0^2 = \frac{k_\sigma \tilde c_d}{\pi^2}  k_1^{d/4}
\ln\frac{1+k_1 }{k_1 }\,(L/L_E)^d\,,
 \eqno(62)
 $$
which should be identified with the Altshuler and Shklovskii
result (32): it gives a relation between  $k_1$ and $k_\sigma$.
The critical point $u_c$ is determined by condition
$Bu_c=(1+u_c)^{1/2}$ following from $\xi=\infty$, and
the first equation (60) should give
$\sigma_0^2/\sigma_P^2=\kappa_0$ for
$u=u_c$. Considering all parameters as
functions of  $k_1$, we have a sequence of relations
$$
k_\sigma = A_d \left[ k_1^{d/4}
\ln\frac{1+k_1 }{k_1 }\right]^{-1}\,,
$$
$$
 B=\frac{2\pi^2
k_1^{1/2}}{(d-2) k_\sigma}\,,
$$
$$
u_c=\frac{1+(1+4B^2)^{1/2}}{2B^2}\,,
 \eqno(63)
$$
$$
\kappa_0=k_1 u_c\ln \frac{1/k_1+1+u_c}{1+u_c} \,,
 $$
where
$$
 A_d=\frac{4\cos(\pi d/4)}{d(1-d/2)} \,,
 \eqno(64)
 $$
and the change of  $k_1$ allows to adjust the correct
$\kappa_0$ value. The actual choice of parameters for
$d=3$ corresponds to   $\kappa_0=0.28$
\cite{11}:
$$
k_1=0.0346\,, \quad k_\sigma = 6.92\,,
$$
$$
B=0.531\,, \quad u_c=4.36\,.
\eqno(65)
$$
The calculated dependence $y=F_\sigma(x)$ is presented
in Fig.\,6,a and compared with the numerical results 
  \cite{11} in Fig.\,6,b.
\begin{figure}
\centerline{\includegraphics[width=2.8 in]{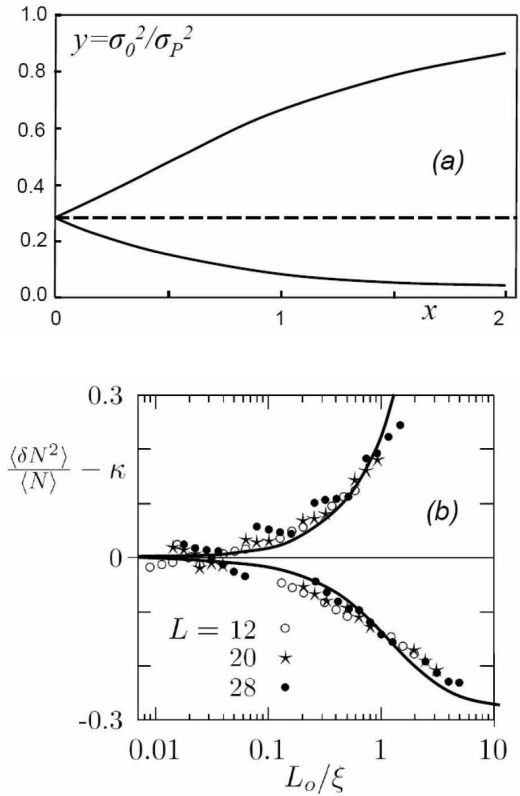}}
\caption{\footnotesize A theoretical dependence for
$y=\sigma_0^2/\sigma_P^2$ as a function of $x=L/\xi s^{1/d}$ (a) and its
comparison with numerical results of the paper \cite{11} (b),
where designation  $L_o=L s^{-1/3}$ was used.} \label{fig6}
\end{figure}

\begin{center}
{\bf 6.2. Scaling for $\sigma^2$ and $A$. }
\end{center}

We have established in Sec.\,2 that $\sigma^2$ and $\sigma_0^2$
coincide in the order of magnitude. The scaling
equations (60) are the same for them, and they differ only by the
choice of parameters. The Poisson value for  $\sigma^2$
is  $s/2$ (see  (12)) and reproduced by the choice
%is achieved with
%is provided by the choice
$k_2=k_\sigma s/2\pi^2$, so  parameter $B$ is two
times less in comparison with (63).  Accepting for
$\sigma^2$ the same behavior in the metallic phase as for
$\sigma_0^2$, we have instead of (63):
$$
k_\sigma = A_d \left[ k_1^{d/4}
\ln\frac{1+k_1 }{k_1 }\right]^{-1}\,,
$$
$$
B=\frac{\pi^2 k_1^{1/2}}{(d-2) k_\sigma}\,,
$$
$$
u_c=\frac{1+(1+4B^2)^{1/2}}{2B^2}\,,
 \eqno(66)
$$
$$
\kappa=\frac{1}{2} k_1 u_c\ln \frac{1/k_1+1+u_c}{1+u_c} \,,
 $$
Parameter $k_1$ is chosen from the critical value
$A_c=1/2\kappa=1.9$ \cite{12} of the scaling variable
$A$ (see (17)), which determines the values of other
parameters:
$$
k_1=0.0366\,, \quad k_\sigma = 6.74\,,
$$
$$
 B=0.280\,, \quad
u_c=13.67\,.
\eqno(67)
$$
Due to relation (17), parameter  $A$ is reversal to
$\sigma^2/\sigma_P^2$  and its scaling is trivially obtained
from Eqs.60. Comparison with the Zharekeshev and Kramer data
\cite{12} is given in Fig.\,7.
\begin{figure}
\centerline{\includegraphics[width=3.0 in]{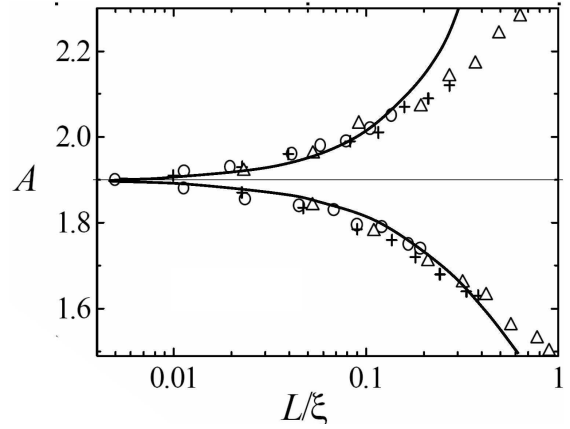}}
\caption{\footnotesize Numerical data by Zharekeshev and
Kramer \cite{12} for the scaling parameter
$A=\sigma_P^2/\sigma^2$ and their comparison with a
theoretical dependence. } \label{fig7}
\end{figure}

\begin{center}
{\bf 6.3. Scaling for   $\alpha(s_0)$. }
\end{center}

The scaling parameter  $\alpha(s_0)$ is also determined by
combination $\sigma^2/\sigma_P^2$, as clear from equation
(16). The latter is valid for  $s_0\gg 1$ and its extrapolation
%to values of the order of unity cannot be reliable; so instead
to values $s_0\sim 1$ cannot be reliable; so instead
of $s_0$ some effective value  $s_{eff}$ should be
used.

Next, one should have in mind that for finite  $s$ the
quantity  $\sigma^2/\sigma_P^2$ does not tend to zero in the
metallic phase. This point can be taken into account, if
the following interpolation formula is accepted  for the function
$F(x)$ in (37)
%one accept for the function  $F(x)$ in (37) the
%following interpolation formula
$$
F(1/x)=1+\tilde c_d x^d\,,
%\eqno(xx)
$$
which provides the correct limits (38); its substitution
into (44) and (59) leads to the change in the second equation
(60):
$$
\pm \left(\frac{L}{s^{1/d}\xi }\right)^{d-2} =
\frac{ (1+u)^{1/2} -B(u-u_0)}{(u-u_0)^{2/d}} \,,\qquad
\eqno(60')
$$
where  $u_0\sim 1/s$. Then $u\to u_0$ in the metallic phase
for  $L\to\infty$, and $\sigma^2/\sigma_P^2$ tends to a finite
value.  If parameters for $\sigma^2$ are chosen in
correspondence with Sec.\,6.2, then the proper choice of
 $s_{eff}$ and $u_0$ allows to provide the correct values of
  $\alpha$ at the critical point and in the metallic region.

Scaling of parameter $\alpha(s_0)$ was studied for  $s_0=2$
in the paper \cite{10} and for  $s_0=0.473$ in the papers
\cite{11}.  These results agree with the theoretical dependence,
if the choice  $s_{eff}=2.22$, $u_0=8.67$ is made in the first
case (note that  $s_{eff}$ is close to $s_0$)  and
$s_{eff}=2.99$, $u_0=10.2$ in the second case (Fig.\,8).  A small
shift along the horizontal axis in Fig.\,8,a corresponds to
addition of the positive value $L_0$ to the length $L$, in
agreement with  corrections to scaling (Sec.\,6.4). It should be
noted, that finiteness of  $u_0$ practically does not affect
the results beyond the metallic region.
\begin{figure}
\centerline{\includegraphics[width=3.0 in]{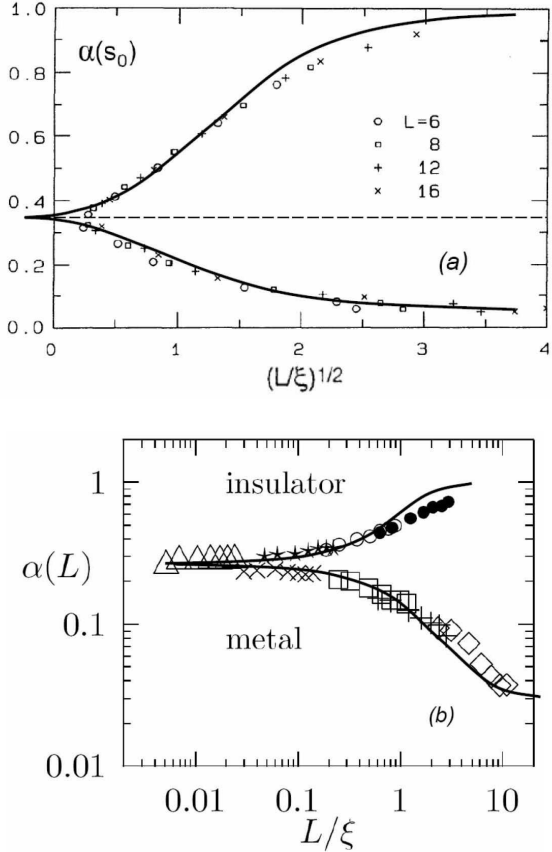}}
\caption{\footnotesize A theoretical
dependence of $\alpha(s_0)$ on $L/\xi$ and its comparison  with 
numerical data of papers
\cite{10} (a) and \cite{11} (b). Values $s_{eff}=2.22$,
 $u_0=8.67$ were used in the first case and $s_{eff}=2.99$,
 $u_0=10.2$ in the second one. } \label{fig8}
 \end{figure}

\begin{center}
{\bf 6.4. Critical Behavior and Corrections to Scaling }
\end{center}

The simplest way to extract the critical behavior from
scaling relations of type  (7) is based on the possibility
to rewrite them in the form ($\tau$ is a distance
to the transition)
$$
\alpha =
F\left(\frac{L^{1/\nu}}{\xi^{1/\nu}} \right) =
F\left( \tau L^{1/\nu} \right) \approx
\alpha_c +C \tau L^{1/\nu}
\eqno(68)
$$
and expand regularly over  $\tau$, which is
possible due to the absence of phase transitions in finite
systems. Then the derivative over $\tau$ behaves as
$L^{1/\nu}$ and immediately determines the critical exponent
$\nu$ of the correlation length  $\xi$.

Such procedure is certainly correct, if  relation  (7)
is exact. In fact, it is not exact due to
existence of scaling corrections. To analyze the latter,
consider a decomposition of the sum over ${\bf q}$ in
(49) suggested in \cite{2}:
$$
\frac{1}{ L^{d}}
\,\frac{1}{ m^2} + \frac{1}{ L^{d}} \sum\limits_{
\begin{array}{c} { \scriptstyle {\rm q}\ne 0 } \\
{\scriptstyle |{\rm q}|<\Lambda  } \end{array}
 }
\left( \,\frac{1}{m^2 + q^2} -\,\frac{1}{ q^2}\right)\,+
$$
$$
+\,\frac{1}{L^{d}} \sum\limits_{
\begin{array}{c} { \scriptstyle {\rm q}\ne 0 } \\
{\scriptstyle |{\rm q}|<\Lambda  } \end{array}
}
 \frac{1}{q^2 } \, \,\equiv \,
I_1(m)+I_2(m)+I_3(0)
\,,
\eqno(69)
$$
where we separated the term with ${\bf q}=0$, and rearrange the
rest sum by subtraction and addition of the same sum with  $m=0$.
Setting  ${\bf q}=2\pi{\bf s}/L$ in the second term  $I_2(m)$,
we can represent it in the form
$$
I_2(m)=L^{2-d} H_0(mL) +O(m^2 \Lambda^{d-4})  \,,
\eqno(70)
$$
where the first term corresponds to the limit
$\Lambda\to\infty$ ($H_0(z)$ is a certain function), and the
second gives a correction, related with finiteness of
$\Lambda$. The third term in (69) can be estimated by
the change of summation by integration with
restriction $|q|\agt 1/L$
$$
I_3(0) = \Lambda^{d-2}  \left[ b_0
+ b_1 \left(\frac{a}{L} \right)^{d-2} \right]   \,.
\eqno(71)
$$
Then, setting $\tau ={\cal E}^2/W^2-b_0 \Lambda^{d-2}$, one
has deviation of the quantity $y=\xi_{0D}/L$ from its critical
value:
$$
\frac{\xi_{0D}}{L} - y_c =
C \left(\frac{L}{a} \right)^{d-2}
\left[\tau + O \left( \frac{a^2}{\xi_{0D}^{2}} \right) \right]
+ O \left(\frac{a}{L} \right) .
\eqno(72)
$$
Differentiating over  $\tau$ and resolving for
$(\xi_{0D})'_\tau$ in the iterational manner, one has
$$
\left(\frac{\xi_{0D}}{L}\right)'_\tau = C_0 L^{d-2}
+ C_1 L^{2d-6}   \,.
\eqno(73)
$$
In three dimensions, the main correction to scaling reduces
to a constant, and  for small $\tau$ we obtain
$$
\frac{\xi_{0D}}{L}-y_c= C_0 \tau \left( L + L_0\right) \,,
 \eqno(74)
$$
neglecting the terms disappearing at  $L\to\infty$. All scaling
parameters are functions of  $\xi_{0D}/L$ and their deviations
from critical values behave analogously.

Result  (74) was obtained in \cite{1} for other scaling
parameter, while its universality was motivated by considerations
based on the Wilson renormalization group. Since the results
for $L$, lesser than some value $L_{min}$,  always fall out of
the scaling picture and are rightfully neglected by researches,
dependence $L+L_0$ with $L_0>0$ can be interpreted as
$L^{1/\nu}$ with $\nu>1$:  such ambiguity of treatment
was demonstrated in  \cite{1,3} on a lot of examples. The results
for level statistics are illustrated in Fig.\,9.
\begin{figure*}
\centerline{\includegraphics[width=5.0 in]{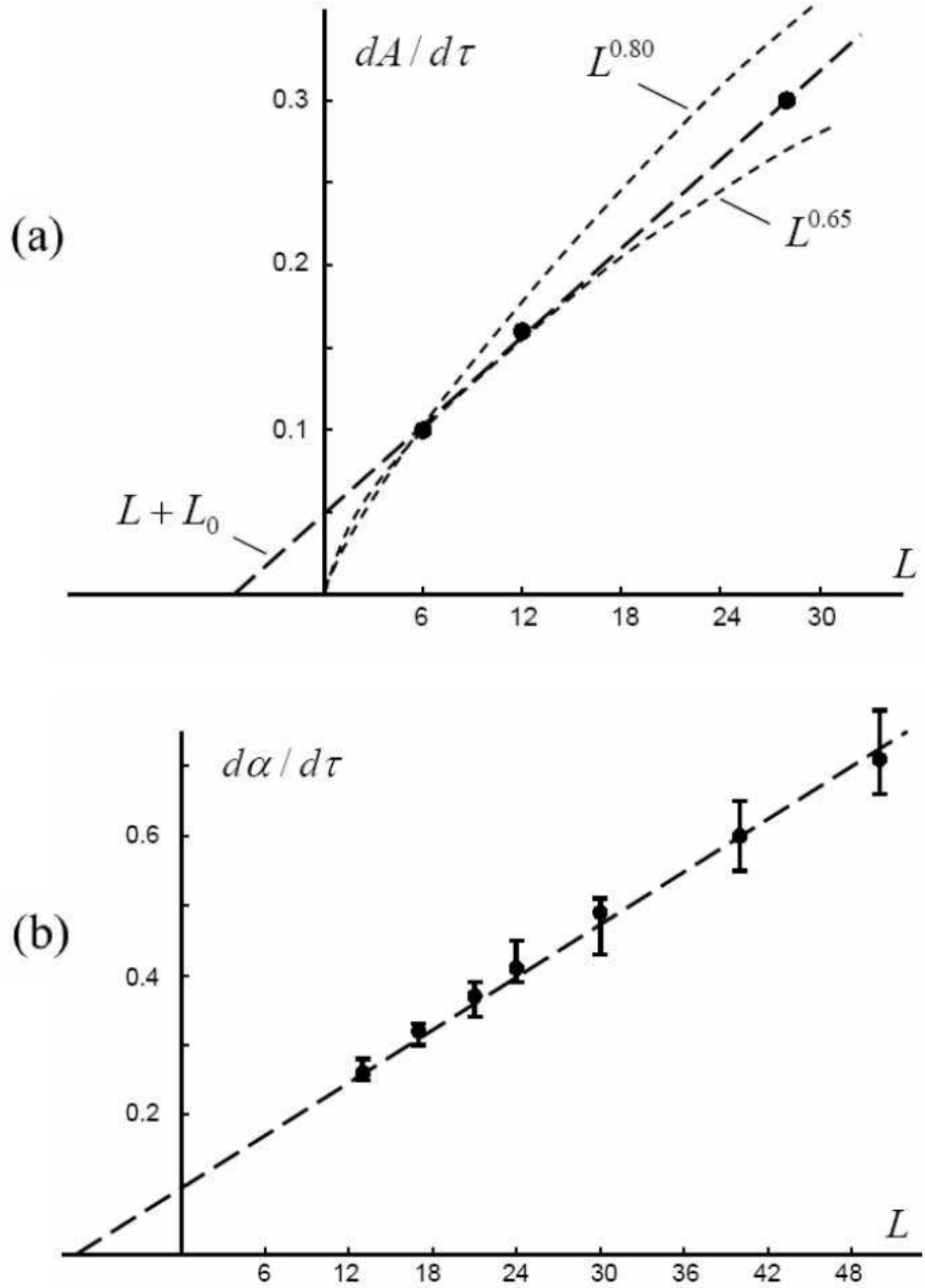}}
\caption{\small Fitting by dependence $'(L+L_0)$
(dashed line) for numerical data, based on
 the level statistics:
 (a) Data by Zharekeshev and Kramer
 \cite{12}. The points correspond to the average
derivatives of the scaling parameter  $A$ (arbitrary units),
determined from Fig.\,4  of \cite{12} in the interval $16<W<17$.
A statistical error related with each point can be estimated
very conservatively (see Table in \cite{28}) due to the irregular
character of curves given in the indicated figure; uncertainty
allowed by the authors themselves corresponds to the gap between
dependencies  $L^{0.80}$ and $L^{0.65}$, determining the upper and
lower bound of the result for the critical exponent, $\nu=1.40\pm
0.15$.
(b) Data obtained by Schreiber' group \cite{15}; the points
correspond to the derivative of the scaling parameter  $\alpha$
(arbitrary units) determined by the slope of solid lines
in the inset of Fig.\,3 in \cite{15};
their uncertainty is obtained by variation of the slope
allowed by the size of experimental points.}
\label{fig9}
\end{figure*}

\begin{center}
{\bf 7. Two-Dimensional Case}
\end{center}

In two dimensions, the power-law function
in the second equation  (51) is replaced by a
logarithmic one  \cite{2},
$$
- c_2 \ln \left(L/\xi\right) =
\frac{p}{z^2}+ H \left(z \right) \,,
\quad c_2=1/2\pi
\eqno(75)
$$
where asymptotics $H(z)=-c_2 \ln z$ is sufficient for large
$p$. Setting as previously
$p=\left[k_1 s^2+ k_2 F(1/z)\right]^{1/2}$,
accepting  $k_2=2k_\sigma s/\pi^2$  in
correspondence with the Poisson condition for  $\sigma^2$
(Sec.\,6.2) and excluding  $z$,  one comes to the following
equation
$$
- \ln\left(\frac{L}{s^{1/2}\xi }\right) = B
\frac{\,\, (1+u)^{1/2} }{u-u_0} - \ln \sqrt{u-u_0}
\eqno(76)
$$
instead of the second equation (60). Here
 $u_0\sim 1/s$ takes into account the finiteness of $s$
in accordance with Sec.\,6.3,
$$
B= \frac{k_\sigma}{\pi^2 k_1^{1/2}}
 =\left[ \pi k_1 \ln \frac{1+k_1}{k_1}  \right]^{-1} \,,
 \eqno(77)
$$
and the relation between  $k_\sigma$ and $k_1$ is used,
obtained from the correspondence with  (32). Parameter
$k_1$ remains free and can be used as a fitting one. For
large  $s$,  the scaling relation (61) remains valid.

In two dimensions, the more complicated scaling parameter was
used \cite{13},
$$
\gamma(s_0)={\cal N}^{-1} \int_0^{s_0} \, \left[
\tilde I(s) - \tilde I_P(s)  \right] ds \,=
$$
$$=
{\cal N}^{-1} \int_{s_0}^\infty \, \left[
 I(s) -  I_P(s)  \right] ds   \,,
\eqno(78)
$$
where the normalization factor  ${\cal N}$ is fixed by the
condition that $\gamma(s_0)=1$ for   $I(s)=I_W(s)$.  The second
equality in (78) follows from the first one due to relation
$I(s)=1-\tilde I(s)$ and the normalization of $I(s)$:
$$
\int_0^\infty \, I(s) ds \,= \int_0^\infty \,s P(s) ds
=\langle s \rangle =1    \,.
\eqno(79)
$$
For large  $s_0$, the second integral in Eq.78 can be estimated
setting $I(s)\sim \exp\{-s \sigma_P^2/\sigma^2\}$
(see (11)) and accepting   $\sigma_P^2/\sigma^2$  to be
practically constant,
$$
\gamma(s_0)= 1- \sigma^2/\sigma_P^2
\exp\left\{-s_0\frac{\sigma_P^2-\sigma^2}{\sigma^2}
 \right\}          \,,
\eqno(80)
$$
so $\gamma(s_0)$ is determined by the quantity
$\sigma^2/\sigma_P^2$.

In  paper \cite{13}, the following dependence was
empirically established for large  $L/\xi$:
$$
\gamma(s_0)\sim \sigma^2/\sigma_P^2-1 \sim (\xi/L)          \,.
\eqno(81)
$$
Such dependence does not take place in the present theory:
it is clear from (76,\,60) that $\gamma(s_0)\sim 1/u$,
$u\sim (L/\xi)^2$  and  behavior  $(\xi/L)^2$ is realized
instead of (81). However, such law is valid practically
for exponentially large values of $L/\xi$, while the numerical
data are satisfactorily fitted for $k_1=0.002$ (Fig.\,10)
(a small value of $k_1$ is not surprising, since it was small
in the $3D$ case). The reason for it is as follows: for small
$k_1$, the large values of $u$ and $x=L/\xi s^{1/2}$ are actual, so
the left and right hand sides of (76) change slowly and can be
linearized near some points $u_c$ and $x_c$.  A freedom in
choice of the common scale of  $\xi$ allows to
compensate the zero term of the linear dependence and
provide  proportionality  $u\propto L/\xi$ in the rather
wide  region of parameters. Thereby, dependence  (81) exists
really as an intermediate asymptotics.
\begin{figure}
\centerline{\includegraphics[width=3.3 in]{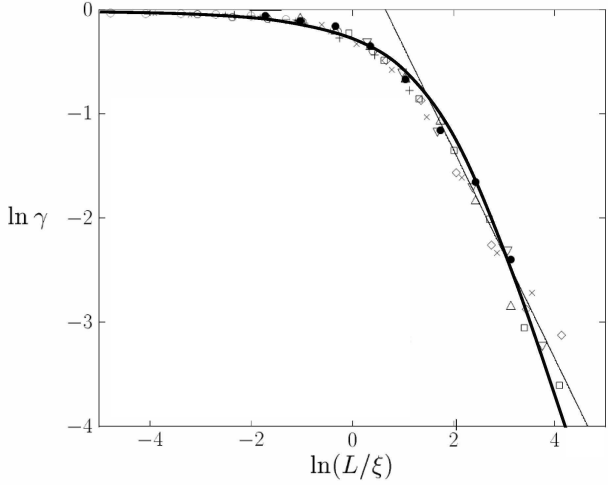}}
\caption{\footnotesize Numerical data of the paper  \cite{13}
for $\gamma(s_0)$ as a function of $L/\xi$ (points) in the $2D$
case and its comparison with the theoretical dependence
for $k_1=0.002$ and $u_0=44$ (thick solid line); value
$s_0=1.25$ was used in
the both cases. The thin solid line corresponds to the law
(81). } \label{fig10}
\end{figure}

\begin{center}
{\bf 8. Higher Dimensions}
\end{center}

\begin{center}
{\bf 8.1. Dimensions $d>4$}
\end{center}

For $d> 4$, one has for the quantity  $I_2(m)$ in  (69)
$$
I_2(m) = - c_{d} m^2 \Lambda^{d-4}\,,\qquad
c_{d} = K_d/(d-4)
\eqno(82)
$$
and the following equation is valid
$$
\pm c_d \left(L/\xi\right)^{2} \left(L/a\right)^{d-4} =
\frac{p}{z^2}-c_d z^{2} \left(L/a\right)^{d-4}
\eqno(83)
$$
instead of the second equation (52).
It is convenient to introduce variables
$$
y=\frac{L}{\xi_{0D}} \left(\frac{L}{a} \right)^{(d-4)/4}\,,\qquad
x=\frac{L}{\xi} \left(\frac{L}{a} \right)^{(d-4)/4}\,,
\eqno(84)
$$
and rewrite  (83)  in the form
$$
\pm c_d x^2 =
\frac{p}{y^2}-c_d y^2\,.
\eqno(85)
$$
Setting as above $p^2=k_1 s^2+k_2 F(1/z)$ and choosing
$k_2 = 2k_\sigma s/\pi^2$ from the Poisson value for the
quantity  $\sigma^2$  (Sec.\,6.2), we have
$$
\pm c_d x^2 = \frac{sk_1^{1/2}(1+u)^{1/2} }{y^2}-c_d y^2\,,
$$
$$
 u=\frac{2k_\sigma}{\pi^2 k_1 s} F(\xi_{0D}/L) \,,
\eqno(86)
$$
where function  $F(1/z)$ is determined  by expression (37) as
previously, but has a different behavior in the actual region of
large $z$,
$$
 F(\xi_{0D}/L) =c_d (L/\xi_{0D})^4 (L/a)^{d-4}=
 c_d y^4  \,.
\eqno(87)
$$
Using (87) and excluding  $y$, we have instead of (60)
$$
\frac{\sigma_0^2}{\sigma_P^2} =k_1 u \ln
 \frac{1+k_1  +k_1 u}{k_1 + k_1 u}\,,
 $$
$$ \pm \frac{x^2}{s^{1/2}} = \frac{
(1+u)^{1/2} -Bu}{u^{1/2}} \,,
\eqno(88)
$$
where $B=\pi^2 k_1^{1/2}/2k_\sigma$.
In the metallic phase, equations (88) give
$$
\sigma^2= \frac{k_1 k_\sigma c_d}{\pi^2} (L/L_E)^4 (L/a)^{d-4}
\ln  \frac{1+k_1 }{k_1 }\,,
 \eqno(89)
$$
which should be identified with the result for the Altshuler
and Shklovskii regime
$$
\sigma^2= \frac{ c_d}{\pi^2} (L/L_E)^4 (L/a)^{d-4}\,,
 \eqno(90)
$$
which follows from (31), but does not coincide with (32).
For a choice of parameters, the relations are valid
$$
k_\sigma = \left[ k_1
\ln\frac{1+k_1 }{k_1 }\right]^{-1}\,, \qquad
B=\frac{\pi^2 k_1^{1/2}}{2 k_\sigma}\,,
 \eqno(91)
 $$
etc., coinciding with (66) for $d=4$.

Equations (88) define in the parametric form the
following scaling relation
$$
\frac{\sigma^2}{\sigma_P^2}
 =F_\sigma \left(\frac{x}{s^{1/4} }\right) \,,  \qquad
 x=\frac{L}{\xi} \left(\frac{L}{a} \right)^{(d-4)/4}\,,
\eqno(92)
$$
which is different from (61) and contains the atomic scale  $a$.
The dependence on  $x\propto L^{d/4}$ instead of $L$ reduces
to the change of the common scale in the logarithmic coordinates,
so a construction of  scaling curves can be produced
in exactly the same manner as in three dimensions;
however, their interpretation should be different and correspond
to (92).

Let emphasize, that in higher dimensions the general form
of the scaling dependence is
$$
\frac{\sigma^2}{\sigma_P^2}
 =F \left(\,\frac{L}{\xi}\,,\, \frac{L}{a}\,     \right)\,,
 $$
since the atomic scale  $a$  cannot be excluded from results due to
nonrenormalizability of theory \cite{25}. At the critical point,
the argument  $L/\xi$ turns to zero, but a dependence on  $L/a$
preserves in the general case: so the scaling parameters
of the standard algorithms are usually  not stationary at the
critical point \cite{1,2}.  Absence of such  a dependence for the
quantity $\sigma^2/\sigma_P^2$ (evident from  (92))
is a nontrivial result of the theory, which agrees with the
existence of the stationary distribution of levels
established in numerical experiments \cite{14}. It should be noted,
that existence of the "spectral rigidity" $\kappa_0$
was related in \cite{18} with constancy of the conductance $g_L$
in the critical point. In higher dimensions, the spectral
rigidity still exists, though $g_L$ is already not constant
\cite{2}.

\begin{center}
{\bf 8.2. Four-Dimensional Case.}
\end{center}

In four dimensions, we have for the quantity  $I_2(m)$
in (69)
$$
I_2(m) = - c_{4} m^2 \ln\frac{\Lambda}{m}
+O(m^4/\Lambda^2)\,,\quad c_{4} = K_4\,
\eqno(93)
$$
and  come to the following equation  instead of (83)
$$
\pm c_4 \left(\frac{L}{\xi}\right)^{2}
\ln\frac{\xi}{a}
= \frac{p}{z^2}-c_4 z^{2} \ln\frac{\xi_{0D}}{a} \,,
\eqno(94)
$$
which in variables
$$
y=\frac{L}{\xi_{0D}} \left(\ln\frac{L}{a}\right)^{1/4}\,,\qquad
x=\frac{L}{\xi} \frac{\left[\ln(\xi/a) \right]^{1/2}}
{\left[\ln(L/a) \right]^{1/4}}\,
\eqno(95)
$$
coincides with (85). Analogously, equation (86) is obtained
with a different behavior of  function $F(1/z)$ at large $z$
$$
F(\xi_{0D}/L) =c_4 (L/\xi_{0D})^4 \ln(\xi_{0D}/a)
\approx c_4  y^4  \,,
\eqno(96)
$$
where we make use of the estimate  $L\sim \xi_{0D} \gg
a$ valid in the critical region.
As a result, equation  (88)  is obtained with a different
definition of  $x$ and the scaling relation holds
$$
\frac{\sigma^2}{\sigma_P^2}
 =F_\sigma \left(\frac{x}{s^{1/4} }\right) \,,  \qquad
x=\frac{L}{\xi} \frac{\left[\ln(\xi/a) \right]^{1/2}}
{\left[\ln(L/a) \right]^{1/4}}\,
\eqno(97)
$$
instead of (92). The usual scaling constructions are possible,
if the quantity $\sigma^2/\sigma_P^2$ is considered as a
function of the "modified length"  $\mu(L)=L [\ln (L/a)]^{-1/4}$.

In the metallic phase, equations (88) give
$$
\sigma^2= \frac{k_1 k_\sigma c_4}{\pi^2} \ln  \frac{1+k_1 }{k_1 }
(L/L_E)^4 \ln (L_E/a) \,,
 \eqno(98)
$$
while in the Altshuler and Shklovskii regime
$$
\sigma^2= \frac{ c_4}{\pi^2} (L/L_E)^4 \ln (L_E/a)\,,
 \eqno(99)
$$
so the previous relations (91) are valid for a choice of
parameters. The actual choice corresponds to  value
$A_c=1/2\kappa=1.4$ \cite{14}:
$$
k_1=0.0652\,, \quad k_\sigma = 5.49\,,
$$
$$
 B=0.230\,, \quad u_c=19.9\,.
\eqno(100)
$$

The main correction to scaling is determined by the term
$O(m^4/\Lambda^2)$ in (93), whose presence in  the
second equation (88) gives for $s=1$
$$
b(u-u_c)=\frac{(L/a)^2 [\tau+ c_4
a^4/2\xi_{0D}^4]}{ [\ln(\xi_{0D}/a)]^{1/2}}
\eqno(101)
$$
where we have linearized the right hand side of  (88) near the
critical point. Differentiating over $\tau$ and resolving for
$u'_\tau$ in the iteration manner, one obtains for small  $\tau$
$$
u=u_c+ \frac{\tau}{b}
\left[ \frac{(L/a)^2)}{[\ln(L/a)]^{1/2}} +
\frac{Q}{[\ln(L/a)]^{2}}\right]     \,,
\eqno(102)
$$
where
$$
Q=  \frac{\pi^2 k_1}{4b k_\sigma} =
\frac{4}{[k_1 \ln(1/k_1)]^{1/2}} = 9.35 \,.
\eqno(103)
$$

In four dimensions, another scaling parameter was used
\cite{14}
$$
J_0={\textstyle\frac{1}{2}}\langle s^2 \rangle =
{\textstyle\frac{1}{2}}\int_0^\infty s^2 P(s)\, ds
=\int_0^\infty s I(s) \,ds \,.
\eqno(104)
$$
It can be estimated setting  $I(s)\sim
\exp(-sA)$ with almost constant $A$ and taking the
normalization (79) of  $I(s)$ into account:
$$
J_0 \approx \left.\frac{\sigma^2}{\sigma_P^2}
\right|_{s\sim 1} \,.
\eqno(105)
$$
Such estimate is rather crude, since the integral is
determined by the region  $s\sim 1$, where $A$ is certainly
not  constant. It is more adequately to consider
 $J_0$ as a regular function of
$\sigma^2/\sigma_P^2$, so deviations of these quantities
from the critical values are proportional to each other
$$
J_0-J_{0c} = const \,
\left(\frac{\sigma^2}{\sigma_P^2} - 2\kappa\right) \,.
 \eqno(106)
 $$
The calculated dependence of $y=\sigma^2/\sigma_P^2$ on
$x$ is presented in Fig.\,11. If a finiteness of $s$ is taken into
account, the quantity  $y$ accepts a finite value in the metallic
phase, and two branches of the dependence become approximately
symmetric. From this point of view, the behavior of
the upper branch is more representative.
\begin{figure}
\centerline{\includegraphics[width=2.7 in]{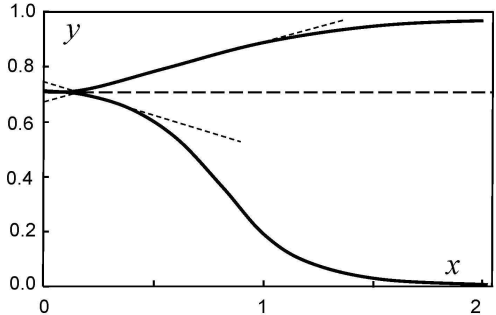}}
\caption{\footnotesize Calculated dependence of
$y=\sigma^2/\sigma_P^2$ on $x$ for $d=4$. The linear
portion in the interval  $0.2 < x < 1$ is clearly seen.  }
\label{fig11} \end{figure}

For the upper branch, one can distinguish  three characteristic
intervals in Fig.\,11: (1) region $x<0.2$ where $y-y_c\sim x^2$,
corresponding to the critical behavior, (2) region  $0.2<x<1$
where the dependence is practically linear, and (3) region of
saturation  $x>1$. The first region corresponds to rather small
values of  $y-y_c$, which are practically unattainable for
numerical experiments due to their restricted
accuracy\,\footnote{\,The narrow critical region is usually
related with existence of small parameters of the Ginzburg number
type.}.  As a result, the observed dependencies (Fig.\,12)
\begin{figure}
\centerline{\includegraphics[width=3.2 in]{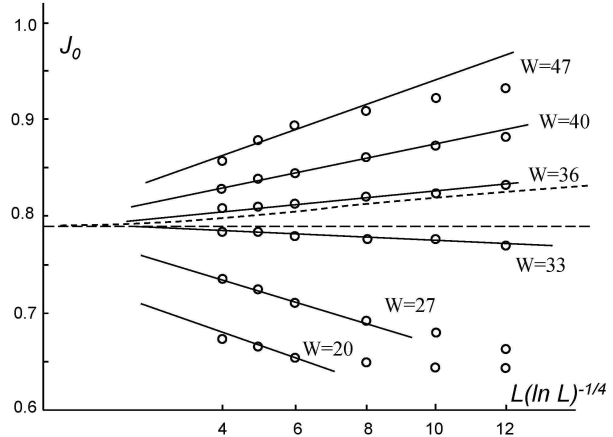}}
\caption{\footnotesize  Numerical data for $J_0$ taken from Fig.4
of the paper  \cite{14} as a function of the modified length
$\mu(L)=L (\ln L)^{-1/4}$ and their fitting by the linear
dependence; the numbers at the horizontal axis marks the
corresponding values of $L$. The dotted line is a theoretical
dependence rescaled in correspondence with the slope of the
linear dependence for  $W=36$; to reach an agreement, the uniform
shift is necessary having the order of  quantity $J_0-J_{0c}$
at $L=4$. } \label{fig12}
\end{figure}
are
close to the linear law $y-y_c=c_1+c_2 x$, and a small $c_1$
value allows to interpret them as  $L^{1/\nu}$ with $\nu\approx
1$ \cite{14}. The ratio of  $c_1$ and $c_2$ is different from
that in the theoretical dependence (Fig.\,11), which can be
explained by corrections to scaling. The main correction is given 
by the second term in the square brackets of Eq.\,102, which is 
a slowly varying (almost constant) function,  becoming
essential for $L/a\alt 3$. Approximately the same uniform shift
 is necessary, in order to provide the correct ratio of  $c_1$
 and $c_2$ (Fig.\,12).

\begin{center}
{\bf 9. Conclusion.}
\end{center}

Accepting validity of self-consistent theory of localization by
Vollhardt and W${\rm {\ddot o}}$lfle, we have derived the
relations of finite-size scaling  for different parameters
characterizing the level statistics. A comparison with the
extensive numerical material shows that on  the level of raw data,
the results of numerical experiments are perfectly compatible with
self-consistent theory, while the opposite statements of the
original papers are related with ambiguity of interpretation and
existence of small parameters of the Ginzburg number type.

Small deviations, which are present in some figures, can be
related with different reasons:

(a) A construction of scaling curves is related with a certain
ambiguity  (see the discussion in \cite{1}). The whole scaling
curve never appears in one experiment but is "measured by pieces".
It is easy to see (Figs.\,6,\,7,\,8,\,10), that the quality of
fitting can be essentially improved, if not the whole curve is
treated but its separate parts.

(b) Existence of scaling corrections (Secs.\,6.4, 8.2)
leads to systematic distortions of the empirical scaling curves.

(c) The exploited above parameters  $k_1,\,k_2,\,k_\sigma$ are in
fact the slowly varying functions and their replacement by
constants is unavoidable approximation related with the absence of
information on these functions.

(d) In some cases, results obtained for  $s_0\gg 1$ are
extrapolated into the region  $s_0\sim 1$.

\noindent
Thereby, reasons (a,b) have a general character, while
(c,d) are specific for the present paper.

In whole, we think it is  possible to say on the realization of
the "minimal program", consisting in elimination of improbably
large (and violating general principles) discrepancies between the
self-consistent theory and numerical experiments. As for the
"maximal program", i.e. testing of the statement that the
Vollhardt and W${\rm {\ddot o}}$lfle theory gives the exact
critical behavior \cite{7,8}, it needs a more detailed analysis of
the existing small deviations and verification of their
significance or insignificance. Such analysis is desirable 
for the initial raw data, and not for empirical scaling
dependencies. It should be noted that in \cite{1,2,3} and the
present paper we have successfully described about 20
dependencies, relating to different quantities and space
dimensions from  2 till  5.

\end{document}